\documentclass{INTERSPEECH2023}
\usepackage{booktabs}
\usepackage{tabularx}
\usepackage{todonotes}
\usepackage{color, colortbl}
\definecolor{Gray}{gray}{0.9}
\usepackage{pifont}
\usepackage{xcolor}
\usepackage[multiple]{footmisc}
\usepackage{bigfoot}
\DeclareNewFootnote{ANote}[fnsymbol]

\interspeechcameraready

\title{Multi-View Multi-Task Representation Learning \\for Mispronunciation Detection}

\name{Yassine El Kheir, Shammur Absar Chowdhury$^*$\thanks{$^{*}$ Corresponding author}, Ahmed Ali}
\address{Qatar Computing Research Institute, HBKU, Doha, Qatar
}
 
\email{
}
\begin{document}

\maketitle
\begin{abstract}

The disparity in phonology between learner's native (L1) and target (L2) language poses a significant challenge for mispronunciation detection and diagnosis (MDD) systems. This challenge is further intensified by lack of annotated L2 data. This paper proposes a novel MDD architecture that exploits multiple `views' of the same input data assisted by auxiliary tasks to learn more distinctive phonetic representation in a low-resource setting. Using the mono- and multilingual encoders, the model learn multiple views of the input, and capture the sound properties across diverse languages and accents. These encoded representations are further enriched by learning articulatory features in a multi-task setup. Our reported results using the L2-ARCTIC data outperformed the SOTA models,
with a phoneme error rate reduction of $11.13\%$ and $8.60\%$ and absolute F1 score increase of $5.89\%$, and $2.49\%$ compared to the single-view mono- and multilingual systems, with a limited L2 dataset.

\end{abstract}


\section{Introduction}

Non-native speakers are often influenced by their mother tongue (L1) when learning a target language (L2). The difference in phonology between the native and target language systems is one of the main factors for pronunciation errors. Computer-aided Pronunciation Training (CAPT) provide personalized and interactive training to help non-native speakers to overcome the negative language transfer effects \cite{article} and improve their pronunciation skills. The mispronunciation detection and diagnosis (MDD) system is a crucial part of the CAPT that detects such pronunciation errors in an L2 learner's speech and provide effective feedback. 

Over the years, various approaches have been investigated, with the majority relying on pre-trained automatic speech recognition (ASR) systems. These methods either (i) identify discrepancies between the aligned ASR output and the reference sequence, or (ii) utilize the log-posterior probability from the ASR to calculate different measures of goodness of pronunciation (GOP) scores, as shown in \cite{GOP_WITT, GOP_1, GOP_2}. 
Deep learning techniques have also been used to train the models, either through an end-to-end approach or by using a cascaded pipeline, where the detection model is trained with GOP features from a pre-trained ASR \cite{JIM, likeJIM_asr, 3M}.

Recently the end-to-end models trained with Connectionist Temporal Classification (CTC) loss have gained popularity due to their promising performance in mispronunciation detection tasks \cite{non_ASR, feng2020sed, yan2020end, wu2021transformer, fu2021full, yan2023peppanet}. 
These CTC-based methods eliminate the need for forced alignment and seamlessly integrate the training pipeline. Furthermore, various of these approaches capitalize on the success of acoustic self-supervised encoders such as wav2vec2.0 \cite{baevski2020wav2vec}. An effective methodology involves the utilization of the CTC decoding techniques to generate sequences of phonemes for MDD \cite{xu2021explore, peng2021study, chen2022alignment}, this approach demonstrates superior outcomes compared to prior research endeavors that solely rely on self-supervised pre-trained models.  




In this study, we introduce a novel MDD framework that disambiguates learners' L1/L2 phonetic representations by utilizing multiple sources of information during training. This information is injected into the model via a multi-view (MV) input representation and multi-task (MT) learning combining primary and auxiliary training objectives. The proposed framework leverages learned representation from monolingual and multilingual pre-trained speech encoders, as an input to the MDD model and fine-tuned it for phoneme sequence recognition along with auxiliary tasks such as classifying different articulatory feature (AFs) sets. The underlying assumption is this ensemble of encoders and auxiliary tasks together provides a unique `view' of the learners' input signal and enriches latent phonetic representation. 
The multi-view representation of the learner's speech enables the model to learn phonetic properties that are either distinct to the target language or shared across multiple languages and accents.  
Our multi-task setup enables the model to capture 
different dimensions of the input signal, it also allows further disentanglement of phonetic properties based on properties of speech production. For instance, the English consonants \textit{/p/}, \textit{/t/}, and \textit{/k/} are all produced with the same manner of articulation (plosive), but they differ in their place of articulation (bilabial, alveolar, and dorsal)

Previous studies \cite{lin2022phoneme, yang2022improving, kheir2023speechblender, fu2022improving} rely on manual frame-level phoneme annotation, canonical phoneme embedding, boundary annotation, pseudo-labeling, and data-augmentation. In contrast, our  framework leverages multiple views of input along with additional learning signals from auxiliary tasks to capture a comprehensive latent phonetic representation that enhance the performance of MDD models in low-resource settings.
\section{Proposed MDD Framework}

\begin{figure} [!ht]
\centering
\includegraphics[width=0.5\textwidth]{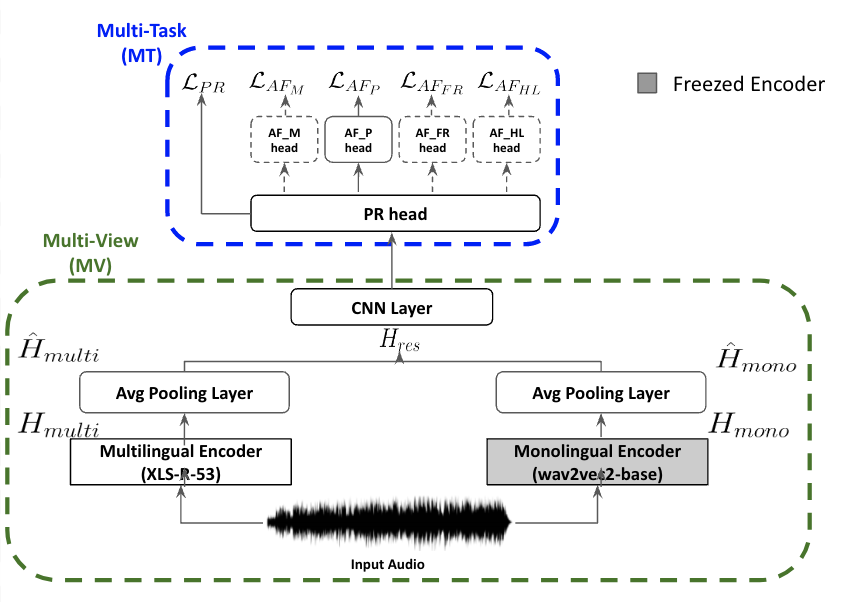}
\caption{Proposed Multi-View Multi-Task Model ($MV_{multi}-MT_{seq}$ Model), with multilingual and monolingual encoders, trained with auxiliary tasks using sequential learning strategy. The weights of the monolingual encoder are frozen.}
\label{fig:model}
\end{figure}

Figure \ref{fig:model} shows the pipeline designed to train the E2E-MDD model using multiple sources of information. 
The model ingests these different views of the input by utilizing representation from different encoders and auxiliary tasks that add additional information and constraints to the MDD model.


\subsection{Multi-View Input Representation Learning}
Given the input raw signal $\mathcal{X} = (x_1, x_2, ..., x_n)$, of n sample length, we first extracted representations from two pre-trained encoders -- monolingual ($\mathcal{H}_{\text{mono}}$) with feature dimension $768$  and multilingual ($\mathcal{H}_{\text{multi}}$) with feature dimension $1024$. We then applied average pooling to both $\mathcal{H}_{\text{mono}}$ and $\mathcal{H}_{\text{multi}}$ to down-sample the feature dimensions to $300$, resulting in $\hat{\mathcal{H}}_{\text{mono}}$ and $\hat{\mathcal{H}}_{\text{multi}}$ respectively.  
Next, we performed element-wise concatenation of $\hat{\mathcal{H}}_{\text{mono}}$ and $\hat{\mathcal{H}}_{\text{multi}}$, resulting in $\mathcal{H}_{\text{res}}$ with feature dimension of $(300,2)$:

\begin{equation}
\mathcal{H}_{\text{res}} = \begin{pmatrix}
\hat{\mathcal{H}}_{\text{multi}} \\
\hat{\mathcal{H}}_{\text{mono}}\
\end{pmatrix} = \begin{pmatrix}
\hat{h}_{\text{multi}_1} & 
\hat{h}_{\text{multi}_2} & ... & \hat{h}_{\text{multi}_{300}} \\
\hat{h}_{\text{mono}_1} & \hat{h}_{\text{mono}_2} & ... & \hat{h}_{\text{mono}_{300}}
\end{pmatrix}
\end{equation}

To further process $\mathcal{H}_{\text{res}}$, we pass it through a simple CNN layer of kernel size $(16,2)$. 
The output multi-view representation is then passed to the phoneme output head ($PR$), which is then trained using the CTC loss ($\mathcal{L}_{PR}$), calculated by comparing the predicted L2 speech phoneme sequence $\hat{\mathcal{Y}} = (\hat{y}_1, \hat{y}_2, ..., \hat{y}_m)$ with the human-labeled phoneme sequences $\mathcal{Y} = (y_1, y_2, ..., y_l)$.

\subsection{Auxiliary Tasks for Multi-Task Learning}
\label{sec:sql}


We enrich the MDD model by adding the auxiliary $\tau$ tasks in multi-task setup. These auxiliary tasks add depth to the phonetic representation, disentangling them by encoding information about speech production properties. We chose to classify quantized levels of four articulatory feature sets, as described in Section \ref{sec:aux}, using 4 individual auxiliary tasks heads, $\tau$ = \{$AF_M$, $AF_P$, $AF_{HL}$, and $AF_{FR}$\}.
These auxiliary task heads ($AF_*$) are simple feed-forward layers, which take the phoneme embeddings from the $PR$ head as input and output sequence of quantized class labels of that task.   
We opt for CTC loss $\mathcal{L}_{AF_{*}}$ to train the auxiliary objective.


We combined the auxiliary loss $\mathcal{L}_{\tau}$ to the main objective loss $\mathcal{L}_{PR}$, based on the training strategy we followed. 
We opt for two different training strategies: (a) combining all the auxiliary loss functions from the beginning of the training 
$\mathcal{L}_{All} = \mathcal{L}_{PR} + \mathcal{L}_{AF_{M}} + \mathcal{L}_{AF_{P}} + \mathcal{L}_{AF_{HL}} + \mathcal{L}_{AF_{FB}}$
;

\noindent(b) gradually adding one auxiliary loss at a time in a specific order, after a warm-up period (

$\mathcal{L}_{Seq} = \mathcal{L}_{PR} + \mathcal{L}_{AF_{*}}$

).
\paragraph*{Sequencial Learning Strategy} 
We designed a scheduler that switches the auxiliary tasks at regular intervals during training to train on a sequence of tasks following a curriculum \cite{bengio2009curriculum}. During the initial warm-up steps, the model learns with only a single task -- i.e. the main objective loss, $\mathcal{L}_{PR}$, we then add one scheduled auxiliary task at a time. These auxiliary tasks are switched at $I$ intervals\footnote{ Note only one auxiliary task is jointly optimized with the main loss at a time.} allowing the model to build upon previously learned representation. This strategy allows the model to learn different properties of sounds without overfitting to any certain properties.

\section{Experimental Setup}

\subsection{L2-ARCTIC corpus}

L2-ARCTIC \cite{L2-ARCTIC} is a non-native English corpus publicly available for research in voice conversion, accent conversion, and mispronunciation detection. It consists of utterances from non-native English speakers, including $12$ males and $12$ females, with L1 languages such as Hindi, Korean, Spanish, Arabic, Vietnamese, and Chinese. The corpus is manually annotated by experts and includes instances of mispronunciation and non-native accents, providing a valuable resource for studying the characteristics of non-native speech. 
We used the common L2-ARCTIC dataset split with $6$ speakers for the test set, $12$ speakers as training, and the remaining $6$ dev sets. For training purposes we mapped the L2-ARCTIC phone set to the mapping table from \cite{lee1989speaker}. 


\subsection{Pre-trained Speech Encoders}
\paragraph*{Monolingual Encoder:} Wav2vec2-base\footnote{https://huggingface.co/facebook/wav2vec2-base} comprises Convolutional Neural Network (CNN) and Transformer layers. The CNNs serve as a feature extractor that converts the input audio waveform $X$ into a latent representation $Z$. Prior to being fed into the Transformer layers, $Z$ undergoes random masking, where a certain portion is masked. The Transformer layers contextualize $Z$ and produce the contextualized representation $C$. The unmasked latent representation $Z$ is further discretized into $Q$ using a learnable codebook. During pre-training, Wav2vec2.0-base is trained using Contrastive Loss only on English Data, which aims to differentiate the true underlying discretized representation $q^+$ for each masked time step $t$ from those at other masked positions ($q^-$), based on the contextualized representation $c_t$. The complete Semi-Supervised Learning (SSL) loss is a weighted sum of the Contrastive Loss and a codebook diversity loss \cite{baevski2020wav2vec}.

\paragraph*{Multilingual Encoder:}

The XLS-R-53\footnote{https://huggingface.co/facebook/wav2vec2-large-xlsr-53} \cite{conneau2020unsupervised} model is a pre-trained wav2vec2.0 \cite{baevski2020wav2vec} model that consists of 53 languages and uses 56,000 hours of speech data for pre-training. The multilingual pre-trained XLS-R-53 model also follows the same architecture as the wav2vec2.0 model. It uses a CNN-based encoder network to encode the raw audio sample and a transformer-based context network to build context representations over the entire latent speech representation. The encoder network consists of $7$ blocks of temporal convolution layers with $512$ channels, and the convolutions in each block have strides and kernel sizes that compress about $25$ms of $16$kHz audio every $20$ms. The context network consists of $24$ blocks with model dimension $1024$, inner dimension $4096$, and $16$ attention heads. 

\subsection{Auxiliary Tasks}
\label{sec:aux}

We used four auxiliary tasks that classify different articulatory properties into their quantized class labels.

\noindent \textbf{Articulatory Features} are the physical properties of speech sounds that can be used to distinguish the close-pair phoneme from each other. The four articulatory feature sets we used are:
\begin{itemize}
    \item \textbf{Place of articulation} refers to the position of the articulators (the tongue, lips, teeth, and palate) in the vocal tract when a speech sound is produced. 
    \item  \textbf{Manner of articulation} refers to the way in which the airstream is obstructed or modified in the vocal tract when a speech sound is produced.
    \item  \textbf{High-low} refers to the position of the tongue in the vocal tract when a speech sound (vowels) is produced. 
    \item \textbf{Front-back} refers to the position of the tongue in the vocal tract when a speech sound (vowels) is produced. 
\end{itemize}
Each articulatory feature is abstracted into several quantized classes, which indicate distinct behaviors \cite{scharenborg2007towards}. These feature sets and their quantized class values (see Table \ref{tab:afs}) classes are obtained  using human-labeled phoneme sequence present in the dataset and the AF mapping proposed in \cite{scharenborg2007towards}.

\begin{table}[htbp]
  \centering
  \renewcommand{\arraystretch}{1} 
  \scalebox{0.9}{
  \begin{tabular}{lp{3cm}cc} 
    \hline
    \textbf{Auxiliary tasks} & \textbf{Classes} & \textbf{\#} & \textbf{Aux Head} \\
    \hline
    \hline
    Manner &  vowel, stop, fricative, retroflex, approximant, nasal, silence & 7 & $AF_M$\\
    \hline
    Place & bilabial, alveolar, dental, labiodental, velar, nil & 6 & $AF_P$ \\
    \hline
    High-low & low, mid, high, nil, & 4 & $AF_{HL}$\\
    \hline
    Fr-back & front, central, back, nil, & 4 & $AF_{FR}$\\
    \hline
  \end{tabular}}
  \caption{AFs and their respective quantized class values}
  \label{tab:afs}
  \vspace{-0.6cm}
\end{table}


\subsection{Model Training and Parameters}
We jointly trained encoders and task-heads (both primary and auxiliary heads) with backpropagation by optimizing the total (multi-task) loss, computed by averaging the losses of the task heads. For simplicity, we use unweighted average loss. All the model components are optimized using Adam optimiser \cite{kingma2017adam}, for $10,000$ steps, with an initial learning rate of $4 \times 10^{-5}$, and batch size of $32$. 

For the sequential learning setup, we ran the primary task (phoneme sequence recognition) only for the first $2,000$ steps, and then fine-tuned the multi-view setup with added auxiliary tasks switched at an equal interval ($I$ = $2,000$ steps). That means in every $2,000$ steps, we switched the auxiliary task. 
The sequence we followed:
Phoneme recognition ($PR$) $\rightarrow$ $PR$ + Manner ($AF_M$) $\rightarrow$ $PR$ + Place ($AF_P$) $\rightarrow$ $PR$ + high-low ($AF_{HL}$) $\rightarrow$ $PR$ + fr-back ($AF_{FB}$).

\noindent We tuned the hyper-parameters with an early stopping criterion using the development set. We used phoneme error rate (PER) to choose the best model and reported results on the test set.


\subsection{Evaluation}
We followed the proposed hierarchical evaluation structure presented in \cite{evaluation_metrics}. We detected pronunciation errors using the predicted and reference phoneme sequence. For the canonical phonemes, we calculated true acceptance (TA) and false rejection (FR) to evaluate model's efficacy in distinguishing correct pronunciation, while to evaluate mispronunciation detection capability, we opted for false acceptance (FA) and true rejection (TR) measures. We then used these measure to calculate the overall recall ($R=\frac{TR}{TR+FA}$), precision ($P=\frac{TR}{TR+FR}$), F1-score, and phoneme error rate (PER).

\begin{table*}[!ht]
    \centering
    \renewcommand{\arraystretch}{1.0}
    \begin{tabular}{lcccccc}
    \toprule
    \textbf{Models} & \textbf{Encoders} & \textbf{Auxiliary-Tasks} & \textbf{R} & \textbf{P} & \textbf{F1} & \textbf{PER}\\ \hline
    \hline
    $SV_{mono}$ & Wav2vec2-base (w2v-b) & \ding{55} & 53.90\% & 54.94\% & 54.42\% & 15.91\%\\ 
    \rowcolor{blue!20}
    $SV_{multi}$ & XLS-R-53 (w2v-x) & \ding{55} & 58.93\% & 56.76\% & 57.82\% & 15.48\%\\ \hline\hline
    $SV_{mono}-MT$ & w2v-b & \ding{51} &  54.05\% & 58.30\% & 56.10\% & 14.60\%\\ 
    \rowcolor{blue!20}
    $SV_{multi}-MT$ & w2v-x & \ding{51} &  59.17\% & 56.74\% & 57.89\% & 15.46\%\\ \hline\hline
    $MV_{mono}$ &  w2v-b + w2v-x$^+$ & \ding{55} &  54.06\% & 56.87\% & 55.43\% & 15.42\% \\ 
    \rowcolor{blue!20}
    $MV_{multi}$ & w2v-b$^+$ + w2v-x & \ding{55} &   \textbf{59.54\%}  & 58.84\% & 59.19\% & 15.32\%\\ \hline\hline
    $MV_{multi}-MT_{all}$ & w2v-b$^+$ + w2v-x & \ding{51} &  56.63\% & 61.04\% & 58.75\% & 14.85\% \\ 
    \rowcolor{Gray}
    \textbf{Proposed $MV_{multi}-MT_{seq}$} & w2v-b$^+$ + w2v-x & \ding{51} & 59.23\% & \textbf{61.43\%} & \textbf{60.31\%} & \textbf{14.13\%} \\ \hline
    \end{tabular}
    \vspace{0.2cm}
    \caption{Reported precision (P), recall (R), F1 and phoneme error rate (PER) for different experimental settings. '$+$' indicates the encoder weight is frozen. The rows highlighted in blue represent the best result in each setting and row in gray shows the overall best performance.}
    
\label{tab:model}
\end{table*}

\begin{table*}[!ht]
\centering
\renewcommand{\arraystretch}{1.0}
\scalebox{0.9}{
\begin{tabular}{lcccccc} 
\toprule
\textbf{Models} & \textbf{Pre-training/Fine-tuning} & \textbf{L2-ARCTIC} (\#Train/\#Test) & \textbf{R} & \textbf{P} & \textbf{F1} & \textbf{PER}\\ \hline
\hline
GOP \cite{DNN_3} & Librispeech \cite{LIBRISPEECH} / - & - / 6 & 69.97\% & 32.54\% & 44.42\% & -  \\  
CNN-RNN-CTC \cite{non_ASR} & - / TIMIT \cite{garofolo1993timit} & 12 / 6 & 74.78\% & 36.76\% & 49.29\% & - \\  
APL-2 \cite{ye2022approach} & - / TIMIT & 12 / 6 & 54.49\% & 52.79\% & 53.62\% & - \\ 

Wav2vec2+ \\Momentum labeling \cite{yang2022improving} & - / UTD-4Accents \cite{ghorbani2019leveraging} & 18 / 6 & 51.20\% & 60.39\% & 55.42\% & \underline{14.36\%}  \\ 
Text-CTC-ATTN \cite{fu2021full} & - / TIMIT & 12 / 6 & 56.12\% & 56.04\% & 56.08\% & 15.58\% \\  
Peppanet \cite{yan2023peppanet} & - / TIMIT &  15 / 6 & 64.53\% & 51.38\% & 56.81\% & - \\ 
Wav2vec2-960h+ \\momentum labeling \cite{yang2022improving} & Librispeech / UTD-4Accents & 18 / 6 & 54.16\% & 58.30\% & 56.16\% & 14.69\%  \\ 
RNN-T \cite{zhang2023phonetic} & - / TIMIT & 12 / 6 & 57.2\% & 60.1\% & \underline{58.6\%} & 15.73\%  \\  
Joint MDD-Align \cite{lin2022phoneme} & - / TIMIT & 18 / 6 & 53.31\% & 77.12\% & \underline{63.04\%} & - \\  \hline\hline



\rowcolor{Gray}
\textbf{Proposed $MV_{multi}-MT_{Seq}$} & -/- & 12 / 6 & 59.23\% & 61.43\% & 60.31\% & \textbf{14.13\%} \\  \bottomrule

\end{tabular}}
\vspace{0.2cm}
\caption{Reported precision (P), recall (R), F1 and phoneme error rate (PER) of different state-of-the-art models and our proposed $MV_{multi}-MT_{Seq}$ model on mispronunciation detection and phoneme recognition tasks.}
\label{tab:prior_res}
\vspace{-0.5cm}

\end{table*}

\section{Results and Discussion}



Table \ref{tab:model} showed that the proposed multi-view multi-task MDD architecture along with sequential learning strategy ($MV_{multi}-MT_{seq}$) outperformed single-view ($SV_*$) models significantly. 
The $MV_{multi}-MT_{seq}$ achieved a phoneme error reduction of $11.13\%$ and $8.6\%$ compared to $SV_{mono}$ and $SV_{multi}$ respectively. Furthermore, the model exhibits noteworthy improvements in absolute F1-score, with an increase of $5.89\%$, and $2.49\%$ compared to $SV_{mono}$ and $SV_{multi}$, respectively. 

\subsection{Model Ablation Study}
Table \ref{tab:model} reported the performance of the proposed architecture in multi-view and multi-task ($MV_{multi}-MT_{seq}$) settings and showed its efficacy over different design variations. 

\noindent The model architecture variation we considered for ablation study are: 
\begin{enumerate}
    \item Single-View ($SV$) setting where we either use monolingual ($SV_{mono}$) or multilingual ($SV_{multi}$) encoder representation;
    \item Single-View with the auxiliary task in multi-task ($SV-MT$) setting for either monolingual ($SV_{mono}-MT$) or multilingual ($SV_{multi}-MT$) encoder representation.
    \item Multi-View ($MV$) setting where we use both mono- and multilingual representation, but we freeze the weights of one of the encoders.
    Here we evaluate monolingual ($MV_{mono}$) and multilingual ($MV_{multi}$) setup with frozen multilingual and monolingual encoders respectively;
    \item Multi-View along with the auxiliary task ($MV-MT$) setting with either sequential ($MV_{multi}-MT_{seq}$) or all-at-once ($MV_{multi}-MT_{all}$) learning strategies.
\end{enumerate}



\paragraph*{Single-View {\em vs} Multi-View} When comparing the $SV_{*}$ to $MV_{*}$, we observed a significant improvement in F1 score (mono: 1.01\%, multi:1.37\%) and in PER reduction (mono:3\% , multi:1\%) using multi-view representation learning. Indicating $SV_{*} < MV_{*}$\footnote{Here $<$ represent the strength of the model architecture.}.


\paragraph*{Effect of the Auxiliary Tasks:} When benchmarking the $SV_{*}$ with $SV_{*}-MT$, we noticed a significant gain: F1 improvement, by $1.68\%$, and PER reduction, by $8.23\%$ using monolingual encoder with auxiliary tasks. When using the multilingual encoder, we noticed a comparable performance with and with-out the auxiliary tasks. Indicating $SV_{*} < SV-MT_{*}$.

\noindent When comparing the multi-view models ($MV_{*}$ {\em vs} $MV_{*}-MT$), we selected the setting where the monolingual encoder is frozen. This is due to the fact that in all these settings -- $SV_*$, $SV_*-MT$, and $MV_*$, the multilingual encoder has consistently outperformed the monolingual counterpart. 
From the results, we observed the efficacy of $MV_{multi}-MT_{seq}$ over $MV_{multi}$. Indicating $MV_{multi} < MV_{multi}-MT$.    


\paragraph*{Effect of Sequential Learning:} We investigate the influence of two different training strategies -- sequential or all-at-once.
For brevity, we reported the effectiveness of sequential learning only on the best architecture (i.e., $MV_{multi}-MT_{seq}$ {\em vs} $MV_{multi}-MT_{all}$).\footnote{We noticed a similar pattern in other $*-MT$ tasks. Note all the $SV-MT_{*}$ model architectures use sequential learning.} We observed with sequential learning, the model outperforms the all-at-once setting by F1 score:$1.56$\% and PER reduction:$4.84$\%. Indicating $MV_{multi}-MT_{all} < MV_{multi}-MT_{seq}$.



\subsection{Comparison with prior works}

Table \ref{tab:prior_res} showcases the experimental outcomes of various methods for mispronunciation detection. These results highlight two key observations. 
First, it is evident that all the E2E DNN methods outperform the GOP approach significantly, particularly in the F1 score. Additionally, while CNN-RNN-CTC achieves the highest recall value, it exhibits a drawback of lower precision. On the other hand, APL-2, Peppanet, Wav2vec2 with additional momentum labeling, and RNN-T demonstrate strong competitiveness among each other and showcase good performance in mispronunciation detection based on the L2-ARCTIC benchmark. 
Second, our proposed \textbf{$MV_{multi}-MT_{Seq}$} MDD model outperforms the aforementioned models with a significant margin of \textbf{1.7\%} in F1 score, along with a PER reduction of \textbf{1.6\%}. These improvements are achieved using only the limited training data of $12$ L$2$ speakers provided by L2-ARCTIC. We abstain from using additional datasets to pre-train (encoders) or fine-tune the model, to emulate low-resource settings.

Our model demonstrates a balanced trade-off between recall and precision, in comparison to the current state-of-the-art approach.
Unlike our model, Joint MDD-Align, relies on manual frame-level phoneme annotation, canonical text-phoneme embedding, and boundary annotation. Moreover, the reported performance of Joint MDD-Align relies on finetuning their model with additional L1-data TIMIT, whereas we achieve comparable results with only limited L2 data.

\section{Conclusion}
In this study, we proposed a novel multi-view multi-task MDD architecture that (a) leverages different views of input representation from mono- and multilingual encoders; (b) capture different properties of speech production from the auxiliary learning objectives; and (c) learns distinctive and rich phonetic representation in a low-resource setting. The proposed MDD model significantly outperformed single-view models with/without additional auxiliary signals for both mispronunciation detection and phoneme recognition tasks. 
Our empirical results suggest that looking into the input signal from different views can effectively capture both distinct and shared patterns in L1 and L2 phonetic representation with limited observation. In future, we will explore how to harness these multi-view models for detecting supra-segmental errors such as intonation among others. 



\bibliographystyle{IEEEtran} 
\bibliography{mybib}
\end{document}